\documentclass[a4paper,12pt, epsfig]{article}
\usepackage[utf8]{inputenc} 

\def\letter{0}\def\pr{0}
\pdfoutput=1 %per jhep

\usepackage{epsfig}
\usepackage{epstopdf}
\usepackage{textcomp}
\usepackage{graphicx}
\usepackage{ifthen}
\usepackage{ulem}

\usepackage{amsmath}
\usepackage[psamsfonts]{amssymb}
\usepackage{euscript}

\usepackage{latexsym}
\usepackage[arrow,matrix,curve]{xy}

\usepackage[hypertexnames=false]{hyperref}
\hypersetup{
    colorlinks=true,
    linkcolor=blue,
    filecolor=magenta,   
    citecolor=black,   
    urlcolor=cyan,
    }

\newskip\humongous \humongous=0pt plus 1000pt minus 1000pt

\newif\ifdtup

\def\,{\hspace{-.1cm}}
\def\hsp{,\hspace{.7cm}}

\def\fc#1#2 {\frac{n}{q}#1\frac{n}{q}#2}

\renewcommand{\cos}{\textrm{cos}}
\renewcommand{\sin}{\textrm{sin}}

\renewcommand{\sinh}{\textrm{sinh}}
\renewcommand{\cosh}{\textrm{cosh}}
\renewcommand{\tanh}{\textrm{tanh}}
\newcommand{\sech}{\textrm{sech}}

\renewcommand{\theequation}{\arabic{section}.\arabic{equation}}
\renewcommand{\(}{\begin{equation}}
\renewcommand{\)}{end{equation} \vspace{-.05in}\linebreak}

\newcounter{saveeqn}
\newcounter{savealpheqn}

\newcommand{\alpheqn}{\setcounter{saveeqn}{\value{equation}}%
  \stepcounter{saveeqn}\setcounter{equation}{0}%
  \renewcommand{\theequation}{\mbox{\arabic{section}.\arabic{saveeqn}
\alph{equation}}}
  \renewcommand{\)}{\end{equation}}}
\def\part#1{\frac{\partial}{\partial{#1}}}%
\def\group#1{\refstepcounter{equation}\setcounter{saveeqn}
 {\value{equation}}%
  \label{#1}\setcounter{equation}{0}%
\renewcommand{\theequation}{\mbox{\arabic{section}.\arabic{saveeqn}
\alph{equation}}}
  \renewcommand{\)}{\end{equation}}}
\newcommand{\reseteqn}{\setcounter{equation}{\value{saveeqn}}%
  \renewcommand{\theequation}{\arabic{section}.\arabic{equation}}%
  \renewcommand{\)}{\end{equation}}}

\newcommand{\aalpheqn}{\setcounter{saveeqn}{\value{equation}}%
  \stepcounter{saveeqn}\setcounter{equation}{0}%
  \renewcommand{\theequation}{\mbox{
        \Alph{subsection}.\arabic{saveeqn}\alph{equation}}}
   \renewcommand{\)}{\end{equation}}}
\newcommand{\areseteqn}{\setcounter{equation}{\value{saveeqn}}%
  \renewcommand{\theequation}{\Alph{subsection}.\arabic{equation}}%
  \renewcommand{\)}{\end{equation}}}

\renewcommand{\(}{\begin{equation}}
\renewcommand{\)}{\end{equation}}
\newcommand{\ba}{\begin{eqnarray}}
\newcommand{\ea}{\end{eqnarray}}

\renewcommand{\sl}{{\sqrt{\lambda}}}

\newcommand{\cbp}{\mathop{\vtop{\ialign{##\crcr
   $\hfil\displaystyle{}\hfil$\crcr\noalign{\kern-13pt\nointerlineskip}
   \BIG{)}\hskip0pt\crcr\noalign{\kern3pt}}}}}
\newcommand{\pa}{\mathop{\vtop{\ialign{##\crcr

$\hfil\displaystyle{\oplus}\hfil$\crcr\noalign{\kern+1pt\nointerlineskip
}
   \hspace{.08in}$^{\alpha=0}$\hskip6pt\crcr\noalign{\kern3pt}}}}}
\renewcommand{\hsp}{,\hspace{.3in}}
\newcommand{\p}{^\prime}
\newcommand{\pp}{^{\prime\prime}}

\catcode`\@=11
\def\vereq#1#2{\lower3pt\vbox{\baselineskip1.5pt \lineskip1.5pt
\ialign{$\m@th#1\hfill##\hfil$\crcr#2\crcr\sim\crcr}}}
\catcode`\@=12

\renewcommand{\(}{\begin{equation}}
\renewcommand{\)}{\end{equation}}

\def\cG{{\mathcal{G}}}
\def\cH{{\mathcal{H}}}

\usepackage[dvipsnames]{xcolor}

\newcommand{\beas}{\begin{eqnarray*}}
\newcommand{\eeas}{\end{eqnarray*}}

\newcommand{\bquo}{\begin{quote}}
\newcommand{\enqu}{\end{quote}}

\def\lim#1{\stackrel{\rm{lim}}{{}_{#1}}}

    \newcommand{\g}{\mathfrak g}

\def\ch{{\mathcal{H}}}

\def\gs#1{\g^{(#1)}}

\def\ok#1{\omega_{k_{#1}}}

\def\XXint#1#2#3{{\setbox0=\hbox{$#1{#2#3}{\int}$}
     \vcenter{\hbox{$#2#3$}}\kern-.5\wd0}}

\newcommand{\beq}{\begin{equation}}
\newcommand{\eeq}{\end{equation}}
\newcommand{\bea}{\begin{eqnarray}}
\newcommand{\eea}{\end{eqnarray}}

\newskip\humongous \humongous=0pt plus 1000pt minus 1000pt

\newif\ifdtup

\catcode`\@=11

\ifthenelse{\equal{\letter}{0}}{ %se lettere=0, commenta questo

\@addtoreset{equation}{section}
\def\theequation{\arabic{section}.\arabic{equation}}

\def\@normalsize{\@setsize\normalsize{15pt}\xiipt\@xiipt
\abovedisplayskip 14pt plus3pt minus3pt%
\belowdisplayskip \abovedisplayskip
\abovedisplayshortskip \z@ plus3pt%
\belowdisplayshortskip 7pt plus3.5pt minus0pt}

\def\small{\@setsize\small{13.6pt}\xipt\@xipt
\abovedisplayskip 13pt plus3pt minus3pt%
\belowdisplayskip \abovedisplayskip
\abovedisplayshortskip \z@ plus3pt%
\belowdisplayshortskip 7pt plus3.5pt minus0pt
\def\@listi{\parsep 4.5pt plus 2pt minus 1pt
      \itemsep \parsep
      \topsep 9pt plus 3pt minus 3pt}}

\relax

\def\section{\@startsection{section}{1}{\z@}{3.5ex plus 1ex minus  .2ex}{2.3ex plus .2ex}{\large\bf}}

\def\thesection{\arabic{section}}
\def\thesubsection{\arabic{section}.\arabic{subsection}}

\def\appendix{\setcounter{section}{0}
 \def\thesection{Appendix \Alph{section}}
 \def\thesubsection{\Alph{section}.\arabic{subsection}}
 \def\theequation{\Alph{section}.\arabic{equation}}}
\renewcommand{\theequation}{\arabic{section}.\arabic{equation}}

}{
\renewcommand{\theequation}{\arabic{equation}}

} %se letter=0, commenta questo

\begin{document}

\def\thefootnote{\fnsymbol{footnote}}
\def\thetitle{Linearized Q-Ball Perturbations}
\def\auttwo{Hui Liu}
\def\autone{Jarah Evslin}
\def\autthree{Tomasz Roma\'nczukiewicz}
\def\autfour{Yakov Shnir}
\def\autfive{Andrzej Wereszczy\'nski}
\def\autsix{Piotr Ziobro}

\def\affd{Institute  of  Theoretical Physics,  Jagiellonian  University,  Lojasiewicza  11,  Krak\'ow,  Poland}
\def\affb{University of the Chinese Academy of Sciences, YuQuanLu 19A, Beijing 100049, China}
\def\affa{Institute of Modern Physics, NanChangLu 509, Lanzhou 730000, China}
\def\affc{Yerevan Physics Institute, 2 Alikhanyan Brothers St., Yerevan 0036, Armenia}
\def\affe{BLTP JINR, Joliot–Curie St 6, Dubna, Moscow region, 141980, Russia}

%\title{Titolo}

\ifthenelse{\equal{\pr}{1}}{
\title{\thetitle}
\author{\autone}
\author{\auttwo}
\author{\autthree}
\author{\autfour}
\author{\autfive}
\author{\autsix}
\affiliation {\affa}
\affiliation {\affb}
\affiliation {\affc}
\affiliation {\affd}
%pr e uno

}{

\begin{center}
{\large {\bf \thetitle}}

\bigskip

\bigskip

%\catcode`@=11

{\large \noindent  \autone{${}^{1,2}$} \footnote{jarah@impcas.ac.cn} {, \auttwo{${}^{3}$}\footnote{ hui.liu@yerphi.am } %\footnote{katarzyna.slawinska@uj.edu.pl},
, \autthree{${}^{4}$}\footnote{tomasz.romanczukiewicz@uj.edu.pl} %\footnote{andrzej.wereszczynski@uj.edu.pl},
, \autfour{${}^{5}$} \footnote{shnir@theor.jinr.ru}
, \autfive{${}^{4}$} \footnote{andrzej.wereszczynski@uj.edu.pl}
and \autsix{${}^{4}$}\footnote{piotr.ziobro@doctoral.uj.edu.pl}
}}

%{\large \noindent  \autone{${}^{1,2}$} \footnote{jarah@impcas.ac.cn} and \auttwo{${}^{1,2}$} \footnote{guohengyuan@impcas.ac.cn}}

\vskip.7cm

1) \affa\\
2) \affb\\
3) \affc\\
4) \affd\\
5) \affe\\

\end{center}

}

\begin{abstract}
\noindent
Linearized deformations of the thick-walled (low-amplitude) (1+1)-dimensional Q-ball may be decomposed into relativistic modes, which are roughly plane waves, and also long-wavelength corotating and counterrotating Floquet modes.  Each mode oscillates at a pair of mirror frequencies which average to the Q-ball frequency. The corotating modes are those of a breather or oscillon plus a very loosely bound mode.  The counterrotating modes are described by an irrational-level P\"oschl-Teller potential, with two discrete modes which mix with their unbound mirrors, unbinding them and turning them into Feshbach-type quasinormal modes.  Expanding to leading order in the Q-ball amplitude, we present all of these modes in closed form, except for the bound mode which does not exist at leading order.

\end{abstract}

\setcounter{footnote}{0}
\renewcommand{\thefootnote}{\arabic{footnote}}

%\maketitle

\section{Introduction}
Consider a scalar field subjected to a potential $V$ with a local or global minimum.  Generically, $V$ will be, at leading order, quadratic about its minimum.  The second derivative at the minimum is the squared mass of the scalar.  If the subleading correction to the potential is negative, then a field oscillating about the minimum will have a lower frequency than the mass gap, with the frequency lowering further as the amplitude increases.  As the frequency lies below the mass gap, such large fluctuations do not linearly couple to the perturbative radiation field, and so such oscillations may be long lived.  In such a case, one says that the theory has a breather \cite{breather} or oscillon \cite{dhn2,osc75}, depending on whether the oscillation eventually decays \cite{segur}.  If there are instead two scalar fields and the potential is axisymmetric about the minimum, then the composition of two oscillons \cite{swap14,o2q} out of phase yields a field which rotates about the minimum, called a Q-ball \cite{qball68,qball}.

If the amplitude of the oscillon or Q-ball is sufficiently small \cite{spector,small97}, then it only probes a small neighborhood of the minimum \cite{qpert08}.  In this case, often called the thick-wall case, the solution and its linearized perturbations are only sensitive to the mass of the field itself and, at leading order in the amplitude, to the leading nonlinearity in the potential.  Classically, this leading nonlinearity, which we will call $\lambda$ below, is dimensionful and so is not a true parameter of the theory.  Quantum mechanically, $\lambda\hbar$ is dimensionless and so the theory has a single parameter, in which one may perform a perturbative expansion.  In either case,  the leading order behavior of the oscillon or Q-ball and its perturbations is insensitive to the higher order nonlinearities.  In fact, in the case of the breather and oscillon it was observed in Ref.~\cite{noiuniv} that the oscillon's nonrelativistic linearized perturbations are entirely independent of the potential of the theory, and the relativistic perturbations are just plane waves which are also independent of the potential.  That the same is true of the Q-ball was noted in Ref.~\cite{qpert08}.  

In the present note, we will use this observation to systematically study the linearized perturbations of the small amplitude Q-ball in a (1+1)-dimensional classical field theory.  
Perturbations of Q-balls have been studied in Ref.~\cite{qpert08,qpert17,qpert18,qpert24,qpert24b,qpert25}.  However, as a result of our small amplitude expansion, we will obtain analytic results, whereas those in the literature are largely numerical\footnote{The nonrelativistic corotating limit that we will study below yields a similar continuum spectrum to that already observed in the case of the nonrelativistic bright soliton of Ref.~\cite{kovtun} and the oscillon \cite{qosc25}.}.

%Like Ref.~\cite{qpert08}, we study small-amplitude Q-balls in 1+1 dimensions, where the amplitude $\epsilon$ is much smaller than the meson mass $m$, but still large enough that the Q-ball charge is much greater than one quantum, so that the semiclassical expansion may be applied \cite{quantq}.  Such small amplitude $Q$-balls are quite insensitive to higher order polynomial terms in the potential \cite{qpert08}, and so our results apply quite generally. 

Note that in quantum field theory, the amplitude of the Q-ball is quantized \cite{weinberg}, like that of the oscillon \cite{qosc25}.  In that setting, we would be interested in an amplitude $\epsilon$ which is many times the fundamental quantum\footnote{This choice is necessary for the validity of the semiclassical expansion, which connects our quantum state to a classical field theory solution.  At subleading orders in the semiclassical expansion, one obtains a rich phenomenology \cite{quantq,qquant24}.}, and yet much smaller than the mass $m$ of the fundamental meson.

In Sec.~\ref{gensez} we review the Q-ball solution and the general form of its linearized perturbations.  In the case of the low amplitude Q-ball, we find the corotating perturbations in Sec.~\ref{corsez} and the counterrotating perturbations in Sec.~\ref{consez}.  Our results are confirmed numerically in Sec.~\ref{numsez}, where we see that the peaks in the power spectrum of a perturbed Q-ball correspond precisely to the discrete nonzero modes described in the previous sections and that they include the discrete modes found in Refs.~\cite{qpert08,qpert24}. 

\section{Generalities} \label{gensez}

\subsection{The Unperturbed Q-ball}

Consider a (1+1)-dimensional classical field theory with a complex scalar field $\phi(x)$ and its dual momentum $\pi(x)$.  Let them be described by the Hamiltonian
\beq
H=\int dx \ch(x)\hsp
\ch(x)=\pi(x) \pi^*(x) +\partial_x\phi(x)\partial_x\phi^*(x)+V(|\phi^2(x)|).
\eeq

We will consider the potential
\beq
V(|\phi^2(x)|)=m^2 |\phi^2(x)|-\frac{\lambda}{4}|\phi^4(x)|+O(\lambda^{n-1}|\phi^{2n}(x)|)\hsp n>2
\eeq
and expand about $\phi(x)=0$. This can be arranged to be a global minimum if desired by choosing the $O(|\phi^{2n}(x)|$ terms appropriately.  In the quantum theory,  to any order in $\lambda\hbar$, $n$ may be chosen to be large enough so that the $O(|\phi^{2n}(x)|$ terms not appear at that order.  Indeed, these corrections will not appear at the leading order of the small amplitude expansion considered in this note.

The corresponding equation of motion is% (2.4) of Ref.~\cite{qpert24}, in the case $\beta=0$
\beq
(-\partial_t^2+\partial_x^2)\phi(x,t)=\frac{\partial V}{\partial |\phi(x,t)|^2}\phi(x,t)=\left(m^2-\frac{\lambda}{2}|\phi(x,t)|^2\right) \phi(x,t) \label{eom}
\eeq
where in the last expression we have dropped the higher order terms.

The Q-ball is a solution of Eq.~(\ref{eom}) of the form
\beq
\phi(x,t)=f(x)e^{i\Omega t}\label{qb}
\eeq
for some profile function $f(x)$.
 Decomposing the complex field $\phi$ into two real fields $\phi_i$
\beq
\phi(x,t)=\frac{\phi_1(x,t)+i\phi_2(x,t)}{\sqrt{2}}
\eeq
this becomes
\beq
\phi_1(x,t)=\sqrt{2}f(x)\cos(\Omega t)
\hsp
\phi_2(x,t)=\sqrt{2}f(x)\sin(\Omega t). \label{qeq}
\eeq

We will be interested in low-amplitude Q-balls, corresponding to the expansion
\beq
\Omega=\sqrt{m^2-\epsilon^2}\hsp f(x)=\frac{2}{\sl}\epsilon\ \sech(\epsilon x)+O(\epsilon^3).
\eeq
Here the $O(\epsilon^3)$ corrections are determined by the higher order corrections to the potential, and vanish in the absence of such corrections.  More formally, we define the low-amplitude Q-ball as the limit $\epsilon/m\rightarrow 0$, in which for simplicity we hold $m$ fixed.
 
\subsection{Linearized Perturbations}

In terms of the real fields, the equation of motion (\ref{eom}) splits into two equations
\beq
(-\partial_t^2+\partial_x^2-m^2)\phi_1(x,t)+\frac{\lambda}{4}\phi_1^3(x,t)+\frac{\lambda}{4}\phi_1(x,t)\phi_2^2(x,t)=0 \label{e1}
\eeq
and
\beq
(-\partial_t^2+\partial_x^2-m^2)\phi_2(x,t)+\frac{\lambda}{4}\phi_2^3(x,t)+\frac{\lambda}{4}\phi^2_1(x,t)\phi_2(x,t)=0. \label{e2}
\eeq

Perturbations of the pair $(\phi_1,\phi_2)$ of real fields can be written in terms of functions $(\gs1,\gs2)$.  As we are interested in infinitesimal perturbations, which satisfy linearized equations of motion, we will introduce a small scale $\delta$ and work to linear order in $\delta$.  Ultimately we will be interested in real perturbations, as $\phi_1$ and $\phi_2$ are real.  However, as in the familiar case of the vacuum sector perturbations, which are plane waves, it will be convenient to allow $\gs1$ and $\gs2$ to be complex.  After we have described a basis of our perturbations, we will impose reality as a condition on the coefficients in this basis.

More concretely, our basis of perturbations of the Q-ball (\ref{qeq}) may be written
\beq
\phi_1(x,t)=\sqrt{2}f(x)\cos(\Omega t)+\delta\ \gs1(x,t)
\hsp
\phi_2(x,t)=\sqrt{2}f(x)\sin(\Omega t)+\delta\ \gs2(x,t). \label{qeqp}
\eeq
Here the $\gs{i}$ are complex functions and $\delta$ is a dimensionless number that is smaller than any power of $\epsilon/m$.   The reality condition is now simple to state.  The total perturbation must consist of terms of the form $a\gs{i}+a^*\g^{(i)*}$ with some complex coefficient $a$. 

%\blu{
%\beq
%\phi_1(x,t)=\sqrt{2}f(x)\cos(\Omega t)+ \eta \ \delta \phi_1(x,t)
%\hsp
%\phi_2(x,t)=\sqrt{2}f(x)\sin(\Omega t)+ \eta \ \delta \phi_2(x,t). \label{qeqp}
%\eeq
%Here the $\delta \phi_i(x,t)$ are real functions and $\eta$ is a dimensionless real number that is smaller than any power of $\epsilon/m$.
%}\blu{Maybe here better to remind the reader what is $\epsilon$, before here $\epsilon$ only appeared once in the Introduction.}

%\blu{This paragraph can be removed $\rightarrow$ }

%To avoid clutter we have considered a single mode $\g$.  Of course, in applications, one adds superpositions of modes.  In particular, as the $\phi_i$ are real, one must remember that in the final expression for $\phi_i$, for each mode, one must add $a\gs{i}+a^*\g^{(i)*}$ with some complex coefficient $a$ \blu{can $a$ be absorbed in $\gs{i}$? Maybe $a$ is a normalized parameter}.

Inserting (\ref{qeqp}) into (\ref{e1}) and (\ref{e2}), one finds, at linear order in $\delta$ 
\beq
\left[-\partial_t^2\,+\,\partial_x^2\,-\,m^2+\frac{\lambda f^2(x)}{4}\left(4+e^{2i\Omega t}+e^{-2i\Omega t}\right) 
\right]\,\gs1(x,t)-i\frac{\lambda f^2(x)}{4}\left(e^{2i\Omega t}-e^{-2i\Omega t}\right) \gs2(x,t)=0 \label{eq1b}
\eeq
and
\beq
\left[-\partial_t^2\,+\,\partial_x^2\,-\,m^2+\frac{\lambda f^2(x)}{4}\left(4-e^{2i\Omega t}-e^{-2i\Omega t}\right) 
\right]\,\gs2(x,t)-i\frac{\lambda f^2(x)}{4}\left(e^{2i\Omega t}-e^{-2i\Omega t}\right) \gs1(x,t)=0. \label{eq2b}
\eeq
%\blu{where the fact was used that (\ref{qb}) was a solution of (\ref{eom}), and the  perturbative $\delta$ terms were kept up to linear order.}

%\blu{
%Now equations (\ref{e1}) and (\ref{e2}) become respectively
%\beq
%\left[-\partial_t^2\,+\,\partial_x^2\,-\,m^2+\frac{\lambda f^2(x)}{4}\left(4+e^{2i\Omega t}+e^{-2i\Omega t}\right) 
%\right]\,\delta \phi_1(x,t)-i\frac{\lambda f^2(x)}{4}\left(e^{2i\Omega t}-e^{-2i\Omega t}\right) \delta \phi_2(x,t)=0 \label{eq1b}
%\eeq
%and
%\beq
%\left[-\partial_t^2\,+\,\partial_x^2\,-\,m^2+\frac{\lambda f^2(x)}{4}\left(4-e^{2i\Omega t}-e^{-2i\Omega t}\right) 
%\right]\,\delta \phi_2(x,t)-i\frac{\lambda f^2(x)}{4}\left(e^{2i\Omega t}-e^{-2i\Omega t}\right) \delta \phi_1(x,t)=0 \label{eq2b}
%\eeq
%where the fact was used that (\ref{qb}) was a solution of (\ref{eom}), and the  perturbative $\eta$ terms were kept up to linear order.
%}

\subsection{The Ansatz}

Let us try to solve these equations using the Ansatz
%\blu{
%\beq 
%\delta \phi_1(x,t)=a \gs1(x,t) + a^* \g^{(1)*}(x,t)
%\hsp
%\delta \phi_2(x,t)=a\gs2(x,t)+ a^* \g^{(2)*}(x,t) \label{cordphi}
%\eeq 
%where}
\beq
\gs1(x,t)=G(x)e^{i(-\Omega-\omega)t}+H(x)e^{i(\Omega-\omega)t}\hsp
\gs2(x,t)=I(x)e^{i(-\Omega-\omega)t}+J(x)e^{i(\Omega-\omega)t} \label{gdec}
\eeq
where $G(x)$, $H(x)$, $I(x)$ and $J(x)$ are all complex functions. 
Inserting this Ansatz into Eq.~(\ref{eq1b}) and {\it{choosing}} to make the $e^{i(3\Omega-\omega)t}$ terms vanish yields the condition
\beq
J(x)=-iH(x). \label{m1}
\eeq
This same condition implies that the $e^{i(3\Omega-\omega)t}$ terms vanish in Eq.~(\ref{eq2b}).  Similarly, {\it{choosing}} to make the $e^{i(-3\Omega-\omega)t}$ terms vanish, one finds
\beq
I(x)=iG(x). \label{m2}
\eeq
In summary, we have chosen to restrict our attention to perturbations with a single frequency, and this choice has led to the condition
\beq
\gs2(x,t)=iG(x)e^{i(-\Omega-\omega)t}-iH(x)e^{i(\Omega-\omega)t}. \label{cor}
\eeq

\subsection{Interpretation} \label{corsub}

%Before turning to the $e^{i(\Omega-\omega)t}$ terms of Eqs.~(\ref{eq1b}) and (\ref{eq2b}), 
Let us pause to interpret Eq.~(\ref{cor}).  The fields $\phi_1(x,t)$ and $\phi_2(x,t)$ are real.  This means that their perturbations must also be real.  

Of course $\gs1$ and $\gs2$ are complex.  The total corresponding perturbations, for a fixed solution of Eqs.~(\ref{eq1b}) and (\ref{eq2b}), must therefore be
\beq
\delta \phi_1(x,t)=a \gs1(x,t) + a^* \g^{(1)*}(x,t)
\hsp
\delta \phi_2(x,t)=a\gs2(x,t)+ a^* \g^{(2)*}(x,t). 
\eeq
Now, substituting in our Ansatz 
%\bea
%\delta \phi_1(x,t)&=&a 
%G(x)e^{i(-\Omega-\omega)t}+aH(x)e^{i(\Omega-\omega)t}
%+ a^*G^*(x)e^{i(\Omega+\omega)t}+a^*H^*(x)e^{i(-\Omega+\omega)t}\nonumber\\
%&=&2\re{aG(x)}\cos((\Omega+\omega)t)+2\re{aH(x)}\cos((\Omega-\omega)t)\nonumber\\
%&&+2\im{aG(x)}\sin((\Omega+\omega)t)+2\im{aH(x)}\sin((-\Omega+\omega)t)
%\eea
%and
%\bea
%\delta \phi_2(x,t)&=&ia 
%G(x)e^{i(-\Omega-\omega)t}-iaH(x)e^{i(\Omega-\omega)t}
%-i a^*G^*(x)e^{i(\Omega+\omega)t}+ia^*H^*(x)e^{i(-\Omega+\omega)t}\nonumber\\
%&=&2\re{aG(x)}\sin((\Omega+\omega)t)+2\re{aH(x)}\sin((\Omega-\omega)t)\nonumber\\
%&&-2\im{aG(x)}\cos((\Omega+\omega)t)+2\im{aH(x)}\cos((-\Omega+\omega)t).
%\eea
%Assembling these into our complex field $\phi(x,t)$ 
one finds a total perturbation of
%\bea
%\delta \phi(x,t)&=&\frac{\delta \phi_1(x,t)+i\delta \phi_2(x,t)}{\sqrt{2}}\label{sbagl}\\
%&=&\sqrt{2}\re{aG(x)}e^{i(\Omega+\omega)t}+\sqrt{2}\re{aH(x)}e^{i(\Omega-\omega)t}\nonumber\\
%&&-\sqrt{2}i\im{aG(x)}e^{i(\Omega+\omega)t}+\sqrt{2}i\im{aH(x)}e^{i(\Omega-\omega)t}\nonumber\\
%&=&\sqrt{2}a^*G^*(x) e^{i(\Omega+\omega)t}+\sqrt{2}aH(x) e^{i(\Omega-\omega)t}.\nonumber
%\eea
\bea
\delta \phi(x,t)&=&\frac{\delta \phi_1(x,t)+i\delta \phi_2(x,t)}{\sqrt{2}}\label{tpert}\\
&=&\frac{a \gs1(x,t) + a^* \g^{(1)*}(x,t) +i [a\gs2(x,t)+ a^* \g^{(2)*}(x,t)]}{\sqrt{2}}\nonumber\\
&=&\frac{a [\gs1(x,t) + i\gs2(x,t)] + a^* [\gs1(x,t) - i\gs2(x,t)]^*}{\sqrt{2}}\nonumber\\
&=&\sqrt{2} a H(x) e^{i(\Omega-\omega)t} +\sqrt{2} a^* G^*(x) e^{i(\Omega+\omega)t}.\nonumber
\eea 
We see that the rotations of both the $G$ terms and the $H$ terms differ in frequency from the total Q-ball by $\omega$.  They rotate in the same direction as the Q-ball, except for the $H$ terms when $\omega>\Omega$.  %However, we will see below that $H/G$ is proportional to $\epsilon^2/\omega$ and so the $H$ terms are negligible when $\omega\gtrsim\Omega$.  We conclude that our condition (\ref{cor}) describes corotating modes.

Note that one arrives at the same perturbation if one exchanges $(\omega,G,H)\rightarrow (-\omega,H^*,G^*)$ and so our Ansatz is redundant.  This redundancy may be removed if one restricts attention to $\omega\geq 0$.  We will adopt that convention.  Table~\ref{tab} summarizes the names of the modes that will appear below.

\begin{table}
\begin{center}
\begin{tabular}{|l|l|}
\hline
$\omega$ value&Modes\\
\hline\hline
Not Floquet&Broken Boost or Amplitude Shift\\
\hline
$\omega=0$&Zero Mode (Broken Space or Time Translation)\\
\hline
$\omega<m-\Omega$&Bound Mode\\
\hline
$m-\Omega<\omega<m+\Omega$&Half-Bound Mode\\\hline
$m+\Omega>\omega\sim 2\Omega+O(\epsilon^2/m)$&Counterrotating Quasinormal Mode\\\hline
$m+\Omega\leq \omega\sim 2\Omega+O(\epsilon^2/m)$&Counterrotating Continuum Mode\\
\hline
$\omega\geq m+\Omega$&Continuum Mode\\
\hline
\end{tabular}
\end{center}
\caption{The names of the modes at different values of $\omega$} \label{tab}
\end{table}

\subsection{The Master Equation}

Now the $e^{i(\Omega-\omega)t}$ terms of Eqs.~(\ref{eq1b}) and (\ref{eq2b}) are identical
\beq
\left[(\Omega-\omega)^2-m^2+\partial_x^2+\lambda f^2(x)\right]H(x)+\frac{\lambda f^2(x)}{2} G(x)=0 \label{eq1c}
\eeq
as are the $e^{i(-\Omega-\omega)t}$ terms
\beq
\left[(\Omega+\omega)^2-m^2+\partial_x^2+\lambda f^2(x)\right]G(x)+\frac{\lambda f^2(x)}{2} H(x)=0. \label{eq2c}
\eeq

\section{Corotating Modes} \label{corsez}

\subsection{The $\epsilon$ Expansion}

We are interested in Q-balls with low amplitudes, which necessarily are spatially large.  The low amplitude means that one expects that it will have little effect on radiation, in the sense that monochromatic radiation will be well-described by plane waves, except when the radiation has a wavelength of order the width of the Q-ball itself.  That motivates us in the present section to consider such nonrelativistic radiation.

With an eye to the nonrelativistic limit, let us define 
\beq
\omega=\omega_2\epsilon^2.
\eeq
We will take $\omega_2$ to be $\epsilon$-independent, which will see yields modes with wavenumbers of order $O(\epsilon)$.  With $\omega$ assumed to be of order $O(\epsilon^2)$, Eq.~(\ref{tpert}) implies that both components $G$ and $H$ now rotate with approximately the same frequency $\Omega$ as the Q-ball itself, and so we will refer to such modes as corotating.

The nonrelativistic limit corresponds to the leading order in the $\epsilon/m$ expansion, at which these equations reduce to
\beq
\left[-1-2m\omega_2+\partial_{\epsilon x}^2+4\sech^2(\epsilon x)\right]H(x)+2\sech^2(\epsilon x) G(x)=0 \label{eq1d}
\eeq
and 
\beq
\left[-1+2m\omega_2+\partial_{\epsilon x}^2+4\sech^2(\epsilon x)\right]G(x)+2\sech^2(\epsilon x) H(x)=0. \label{eq2d}
\eeq
Note that $\lambda$ has disappeared, leaving the dimensionful parameter $m$.  As a result, we claim that these modes are universal at leading order in our expansion in the amplitude $\epsilon/m$.
These two equations are, in fact, the same eigenvalue equations that describe the modes of the bright soliton of Ref.~\cite{kovtun} and also in the oscillon, where they are given in Eq.~(5.16) of Ref.~\cite{opert24}.  They are solved by \cite{qosc25}
\bea
G_{\epsilon k/m}(x)\,&=&\,\left(\frac{k^2}{m^2}\,-\,\tanh^2(\epsilon x)\,-\,\frac{2ik}{m}\tanh(\epsilon x)\right)e^{-i\epsilon x k/m}\nonumber\\
H_{\epsilon k/m}(x)\,&=&\,\sech^2(\epsilon x)e^{-i\epsilon x k/m}\label{fl} 
\eea
where
\beq\label{omega_2}
G_{\epsilon k/m}^*=G_{-\epsilon k/m}\hsp H_{\epsilon k/m}^*=H_{-\epsilon k/m}\hsp \omega_{\epsilon k/m}=\epsilon^2\omega_{2,k}=\frac{\epsilon^2}{2m^3}(m^2+k^2)
.
\eeq
Here the $\epsilon k/m$ subscript on $G$, $H$ and $\omega$ means that we are referring to a specific solution, indexed by the real number $k$. Asymptotically these solutions are plane waves with wave number $\epsilon k/m$.

Besides these continuum solutions, there are also two zero-mode solutions at $\omega_2=0$.  If $G=H$ then the equations are of P\"oschl-Teller form with level $\sigma=2$, as in the case of the normal modes of the $\phi^4$ double-well model's kink.  The corresponding solution is the shape mode of that model
\beq
G_B(x)=H_B(x)=\sech(\epsilon x)\tanh (\epsilon x).
\eeq
In the present case, it is not a shape mode, but it is the zero mode of the Q-ball corresponding to the broken translation symmetry.

If, on the other hand, $G=-H$ then $G$ satisfies a P\"oschl-Teller equation at $\sigma=1$, similarly to the normal modes of the Sine-Gordon soliton.  The solution is the soliton's translation zero-mode
\beq
G_T(x)=-H_T(x)=i \ \sech(\epsilon x)
\eeq
which in the case of the Q-ball is the zero-mode corresponding to the broken time-translation symmetry.  The factor of $i$ is included to make $\gs1$ and $\gs2$ real, which is convenient for quantization.

There are also two other such discrete modes, which do not satisfy our periodic Ansatz (\ref{gdec}).  The first corresponds to the broken boost symmetry, which is not periodic because a boost, after evolution by one period, leaves a translation.  This mode is simply the linearized boost.  The second is a change in amplitude, which is not periodic with period $\Omega$ because it changes the period.  This second linearized mode is proportional to the solution itself.

Beyond the nonrelativistic limit, when $m\omega/\epsilon^2\gg 1$, the solutions of Eqs.~(\ref{eq1c}) and (\ref{eq2c}) are simply plane waves
\beq
G_k=e^{-ikx}
\hsp
H_k=0\hsp\ok{}=\sqrt{m^2+k^2}-\Omega.
\eeq

In conclusion, for each $\omega$, we find two solutions, corresponding to right and left-moving $\pm k$ unbound normal modes.

\subsection{Another Bound Mode} \label{boundsez}

How reliable is our small $\epsilon$ limit?  Recall that $\epsilon$ has dimensions of mass, and it is only sensible to consider a limit in which a dimensionless quantities are small.  In our case, we have considered $\epsilon/m$ to be small.   

In particular, the wavenumber divided by $\epsilon$, which we have called $k/m$ is not necessarily large.  On the contrary, it is small close to the mass threshold $\omega=m-\Omega$ where $k=0$.  Therefore one might expect that our expansion is not reliable near the threshold.  

How could the expansion fail?  Note that $\epsilon x$ is also dimensionless, and it is large at sufficiently large $|x|$.  Therefore, one expects the modes found above to be unreliable when $|x|\gg 1/\epsilon$.  Of course in that region, the Q-ball solution tends to zero and so the perturbations are either plane waves for continuum modes or exponentially-decaying for bound modes, and so may seem uninteresting.

\begin{figure}
    \centering
    \includegraphics[width=.45\linewidth]{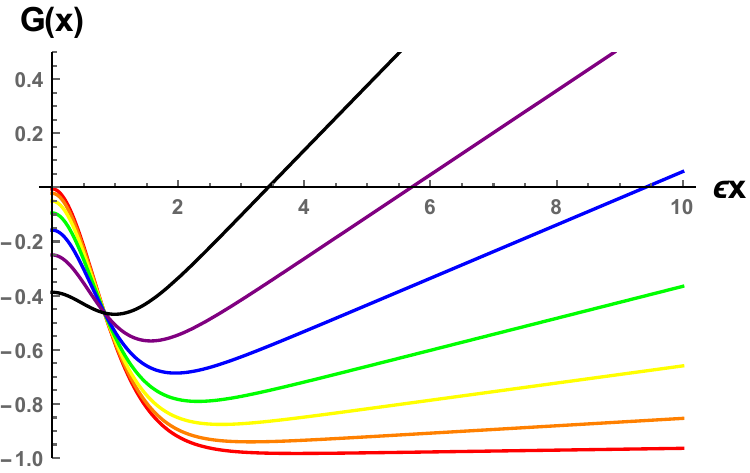}
\includegraphics[width=.42\linewidth]{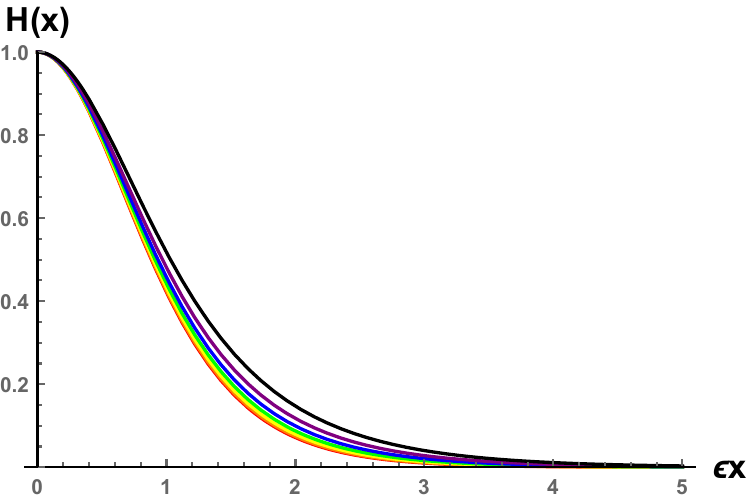}
    \caption{The solutions $G(x)$ (left) and $H(x)$ (right) of Eqs.~(\ref{eq1c}) and (\ref{eq2c}) at the continuum threshold $\Omega+\omega=m$ for $\epsilon$ equal to $0.1$ to $0.7$ in even steps shown in red, orange, yellow, green, blue, purple and ultraviolet respectively.  While the red curve in the $G(x)$ plot is well-approximated by -tanh${}^2(\epsilon x)$ as in Eq.~(\ref{fl}), at large $|x|$ our $\epsilon$ expansion misses the linear rise and the inevitable $x$-intercept.}
    \label{threshfig}
\end{figure}

But what about the threshold solution $k=0$ in Eq.~(\ref{fl})?  At large $|x|$, $G$ tends to $-1$.  The argument above suggests that $\epsilon x$ corrections will be large when $\epsilon |x|\gtrsim 1$, as can be seen in Fig.~\ref{threshfig}.  Here we consider a strict $\phi^4$ potential with no higher order terms, but $\epsilon$ is treated exactly.

Note that for all values of $\epsilon$, at the threshold $k=0$, $G$ increases linearly at large $|x|$ and so it has a zero at positive and negative $x$, although for small $\epsilon$ the zero is at $|x|\sim O(1/\epsilon^3)$.  The existence of this zero implies that there will be a lower energy configuration, which, being below the threshold, will be a bound mode.  As this zero was not evident in our leading order $\epsilon$ expansion, neither is the energy reduction needed to eliminate it, which explains the fact that the bound mode was not found in our analytic treatment above.  Indeed, in the $\epsilon/m\rightarrow 0$ limit, the $\omega_2$ value of this solution is necessarily not fixed.

\begin{figure}
    \centering
    \includegraphics[width=.45\linewidth]{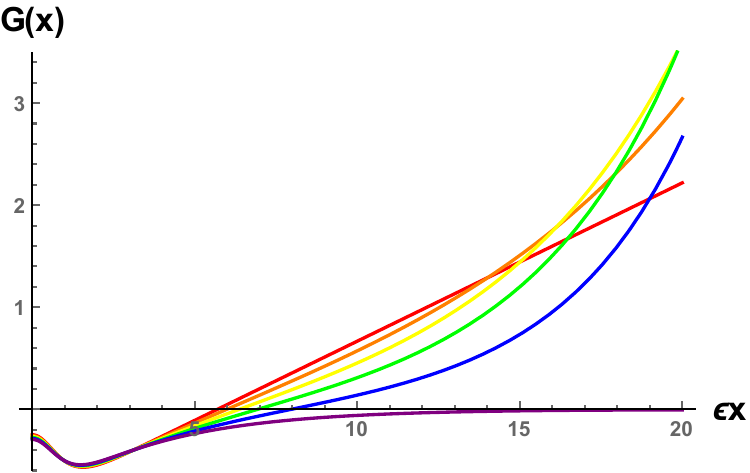}
\includegraphics[width=.42\linewidth]{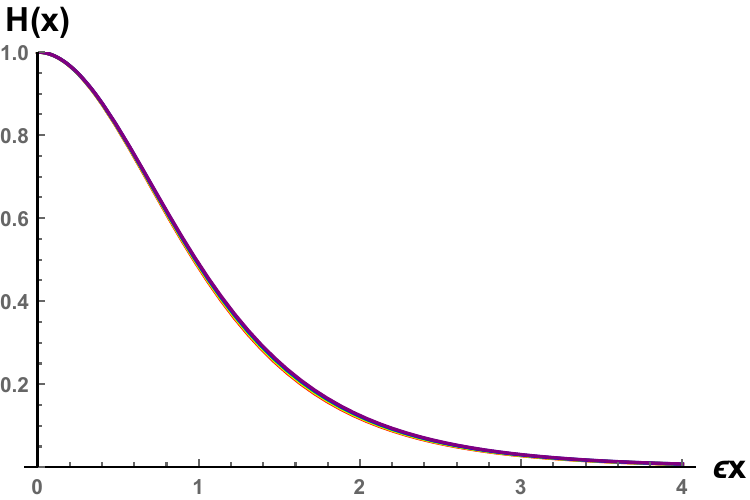}
    \caption{The solutions $G(x)$ (left) and $H(x)$ (right) of Eqs.~(\ref{eq1c}) and (\ref{eq2c}) at $\epsilon=0.6$ at frequencies $\omega$ ranging from the continuum threshold $\omega=m-\Omega$ to the bound state at $\omega=m-\Omega-0.0139$ in even steps shown in red, orange, yellow, green, blue and purple respectively.  The red curve is the threshold mode while the purple curve is the bound mode.  For all solutions in between, $G(x)$ is exponentially divergent.}
    \label{boundfig}
\end{figure}

The solutions interpolating between the threshold and the bound state are shown in Fig.~\ref{boundfig}.  It can be seen that all of these solutions except for the bound state itself have zeros, after which they exponentially diverge.  As expected in the Schrodinger problem, the lower energy bound state is distinguished by having less zeros than higher energy solutions, although the intermediate solutions diverge exponentially and so are not normalizable perturbations.  

The frequency of the bound mode is beneath the threshold by of order $O(\epsilon^6/m^5)$.  This implies that $G$ rises to zero quite slowly, over characteristic lengths of order $O(m^2/\epsilon^3)$.  At small $\epsilon$, this is much larger than the Q-ball itself, and so we conclude that this weakly bound excitation is very delocalized and provides a novel characteristic length scale for the small amplitude Q-ball.  It would be interesting to see the implications of the corresponding extended halo in the quantum theory.

\section{Counterrotating Modes} \label{consez}

In the previous section we investigated the nonrelativistic limit in which $\omega\sim O(\epsilon^2)$ so that both components $G$ and $H$ of a perturbation corotate with the Q-ball.  This is nonrelativistic in the sense that the frequency of the radiation is close to $\Omega$ which is close to the meson mass $m$, and so the wavenumber is small.  However, Eq.~(\ref{tpert}) shows that there is another regime in which the wavenumber is also small, the case in which $\omega\sim 2\Omega$ so that the $H$ component counterrotates with a frequency of roughly $-\Omega$ while the $G$ component rotates with a frequency of roughly $3\Omega$.  In this case $H$ has a wavelength of order the Q-ball size and so again one may expect a large deviation from the plane wave form.

To define such a limit, we define a frequency
\beq
\hat{\omega}=-2\Omega+\omega
\eeq
such that
\beq
\hat\omega_2=\hat{\omega}/\epsilon^2
\eeq
will be kept fixed in our small $\epsilon$ limit, implying $\omega=2\Omega+O(\epsilon^2)$.  Then our Ansatz can be written
\bea
\gs1(x,t)&=&G(x)e^{i(-3\Omega-\hat\omega)t}+H(x)e^{i(-\Omega-\hat\omega)t}\\
\gs2(x,t)&=&iG(x)e^{i(-3\Omega-\hat\omega)t}-iH(x)e^{i(-\Omega-\hat\omega)t}.\nonumber
\eea
%As these will anyway be summed into real combinations, for simplicity we may drop the complex conjugations on the left hand side.  Similarly, we may redefine $G$ and $H$ to be their conjugates, since anyway they solve real equations and so their conjugates will also be solutions.  With these conventions, our Ansatz becomes
%\bea
%\gs1(x,t)&=&G(x)e^{i(-\Omega-\hat\omega)t}+H(x)e^{i(-3\Omega-\hat\omega)t}\\
%\gs2(x,t)&=&-iG(x)e^{i(-\Omega-\hat\omega)t}+iH(x)e^{i(-3\Omega-\hat\omega)t}.\nonumber
%\eea
The deformation of our field is then
\bea\label{g3rot}
\delta \phi(x,t)&=&\frac{a [\gs1(x,t) + i\gs2(x,t)] + a^* [\gs1(x,t) - i\gs2(x,t)]^*}{\sqrt{2}}\\
&=&\sqrt{2} a H(x) e^{i(-\Omega-\hat\omega)t} +\sqrt{2} a^* G^*(x) e^{i(3\Omega+\hat\omega)t}.\nonumber
\eea 
As desired, when $|\hat\omega|\ll m$, the $H$ mode will have a frequency close to the meson mass and so it will have a long wavelength.  It will be counterrotating, and so we will call these counterrotating modes.  However, $G$ will be corotating and also relativistic.

 %For a complex scalar perturbation with a fixed energy, we expect four modes, corresponding to the choices of right moving versus left moving and also to the two real degrees of freedom $\phi_1$ and $\phi_2$.  We have only found two, corresponding to the corotating modes.

%The reason for this is the matching conditions (\ref{m1}) and (\ref{m2}) which fixed $\phi_2$ in terms of $\phi_1$.  Clearly, the perturbations may satisfy any initial condition, and so these conditions are not necessary.

%One choice to obtain the other two solutions is to simply flip the signs in these conditions, so that
%\beq
%J(x)=iH(x)\hsp I(x)=-iG(x). %\label{counc}
%\eeq
%Proceeding as in Subsec.~\ref{corsub} one finds
%\bea
%\delta \phi(x,t)&=&\frac{a [\gs1(x,t) + i\gs2(x,t)] + a^* [\gs1(x,t) - i\gs2(x,t)]^*}{\sqrt{2}}\\
%&=&\sqrt{2} a G(x) e^{i(-\Omega-\omega)t} +\sqrt{2} a^* H^*(x) e^{i(-\Omega+\omega)t}.\nonumber
%\eea 
%We see that such perturbations rotate with the opposite central frequency $-\Omega$ with respect to the Q-ball background.  Therefore we refer to these perturbations as counterrotating modes.  

%Repeating the analysis of Sec.~\ref{corsez} with this choice, the last term on the left hand side of Eqs.~(\ref{eq1c}) and (\ref{eq2c}) vanishes.  Let us normalize $G$ and $H$ so that they are of order $O(\epsilon^0)$.  Then, up to corrections of order $O(\epsilon^4)$, these equations for $G$ and $H$ are decoupled\footnote{The subleading contribution to $G$ will be described in Eq.~(\ref{g3}).}
As $G$ oscillates very quickly, with of order $m/\epsilon$ oscillations over the length of the Q-ball, at leading order in the $\epsilon$ expansion we may ignore its backreaction on $H$.  This leaves us with the following equation for $H$
\beq
%\left[(\Omega-\omega)^2-m^2+\partial_x^2+\lambda f^2(x)\right]H(x)\,=\,O(\epsilon^4)\hsp\,\,\,\,
\left[(\Omega+\hat\omega)^2-m^2+\partial_x^2+\lambda f^2(x)\right]H(x)\,=\,O(\epsilon^4). \label{eq2cn}
\eeq  
This is just Eq.~(\ref{eq2c}) without the last term on the left hand side.  Taking the nonrelativistic limit as above, now with $\hat\omega=\hat\omega_2\epsilon^2$
\beq
%\left[-1-2m\omega_2+\partial_{\epsilon x}^2+4\sech^2(\epsilon x)\right]H(x)=0\hsp
\left[-1+2m\hat\omega_2+\partial_{\epsilon x}^2+4\sech^2(\epsilon x)\right]H(x)=0. \label{eq2d}
\eeq
One recognizes these as exactly solvable P\"oschl-Teller equations in $H(x)$ with level
\beq
\sigma=\frac{-1+\sqrt{17}}{2}.
\eeq
This system has continuum solutions and also two discrete albeit nonzero frequency shape mode solutions.  

%As $G$ and $H$ are decoupled, one may consider modes which contain only $G$ or only $H$.  However, the $G$ term and the $H$ term in the decomposition of $\gs1$ in Eq.~(\ref{gdec}) are related by a shift of $\omega$, and so one may choose to keep either $G$ or $H$.  Of course, switching between $G$ and $H$ will change the sign of $\omega$, and so one must be careful at the end to consider both signs for $\omega$.  However, in the relativistic case $m\omega/\epsilon^2\gg 1$ the solutions are anyway plane waves, and so it is clear that both signs should be kept.  %In the nonrelativistic case, there are no continuum solutions for $H$, and so again this poses no problem.

%With this understood, we will keep only $G$, because unlike $H$, it has apparently discrete solutions for small $\omega$.  Our Ansatz is then
%\beq
%\gs1(x,t)=G(x)e^{i(-\Omega-\omega)t}+O(\epsilon^2)\hsp
%\gs2(x,t)=-iG(x)e^{i(-\Omega-\omega)t} +O(\epsilon^2)
%\eeq
%with the P\"oschl-Teller equation of motion
%\beq
%\left[2m\hat\omega_2-1+\partial_{\epsilon x}^2+4\sech^2(\epsilon x)\right]G(x)=0. \label{eq2e}
%\eeq
Now we see that the equations of motion for the counterrotating modes, like those of the corotating modes, are independent of $\lambda$ at this order and so universal.

Again, as in (\ref{omega_2}) the continuum modes are indexed by
\beq
k=m\sqrt{2m\hat\omega_2-1}
\eeq
where the asymptotic wavenumber is $k\epsilon/m$.  For each $\hat\omega_2>1/(2m)$, this has an even solution
\beq
H_{\epsilon k/m,e}(x)=\cosh^{\sigma+1}(\epsilon x)\ {}_2F_1\left(\frac{\sigma+1-ik/m}{2},\frac{\sigma+1+ik/m}{2};\frac{1}{2};-\sinh^2(\epsilon x)
\right)
\eeq
and an odd solution
\beq
H_{\epsilon k/m,o}(x)=\cosh^{\sigma+1}(\epsilon x) \sinh(\epsilon x)\ {}_2F_1\left(\frac{\sigma+2-ik/m}{2},\frac{\sigma+2+ik/m}{2};\frac{3}{2};-\sinh^2(\epsilon x)\right).
\eeq

%Now however the equations for $e^{i(- 3\Omega-\omega)t}$ are no longer satisfied.  Therefore, there will be corrections to $G$ and $H$ of the form $e^{i(-3\Omega-\omega)t}$, albeit suppressed by a factor of $\epsilon^2$.  More precisely, if we momentarily include the subleading terms in our Ansatz
%\bea
%\gs1(x,t)&=&G(x)e^{i(-\Omega-\omega)t}+G_3(x)e^{i(-3\Omega-\omega)t}+O(\epsilon^4)\\
%\gs2(x,t)&=&-iG(x)e^{i(-\Omega-\omega)t}+iG_3(x)e^{i(-3\Omega-\omega)t} +O(\epsilon^4)\nonumber
%\eea
%where $G_3$ is of order $O(\epsilon^2)$, then the new coefficient $G_3(x)$ must satisfy
Note that $G(x)$ satisfies
\beq
\left((3\Omega+\hat\omega)^2-m^2+\partial_{x}^2\right)G(x)=-\frac{\lambda f^2(x)}{2}H(x)+O(\epsilon^4). \label{g3}
\eeq
$|\hat\omega|\ll m$ in our nonrelativistic limit and so $3\Omega+\hat\omega>m$.  At large $x$, the right hand side vanishes and so we see that $G$ is asymptotically a plane wave.  This means that counterrotating modes are always unbound.

In addition to the continuum modes, there are also two discrete solutions of the P\"oschl-Teller system, where
\beq
2m\hat\omega_{2,e}-1=-(\sigma)^2=\frac{\sqrt{17}-9}{2}\hsp
2m\hat\omega_{2,o}-1=-(\sigma-1)^2=\frac{3\sqrt{17}-13}{2}
\eeq
corresponding to
\beq
\hat\omega_{2,e}=\frac{\sqrt{17}-7}{4m}\sim -\frac{0.719}{m}
\hsp
\hat\omega_{2,o}=\frac{3\sqrt{17}-11}{4m}\sim \frac{0.342}{m}.\label{qnmw}
\eeq
Note that the total frequency
\beq
\Omega+\hat\omega\sim m+\epsilon^2\left(\hat\omega_2-\frac{1}{2m}\right)
\eeq
is less than the mass gap $m$ in both cases, and so these modes would not be able to escape into the bulk were it not for $G$.  However $G$ is asymptotically a plane wave, and so these are in fact Feschbach type quasinormal modes, which do escape.
%\red{Are these related to the two bound modes of Refs.~\cite{qpert08,qpert24}? (They use much larger Q-balls with amplitude $\epsilon=1/2)$.  The frequencies seem different and $\omega_{2,e}<0$, although that could be down to conventions, but it is likely that there are mistakes above.  Anyway, if there are no mistakes above, it would be nice to redo the numerics in Appendix A in Ref.~\cite{qpert24} at a smaller $\epsilon$ to see if the frequencies match those here.}
The even quasinormal mode is
\bea
H_{e}(x)&=&\cosh^{\sigma+1}(\epsilon x)\ {}_2F_1\left(\frac{\sqrt{17}}{2},\frac{1}{2};\frac{1}{2};-\sinh^2(\epsilon x)
\right)\\
&=&\left[\sech(\epsilon x)\right]^{\frac{\sqrt{17}-1}{2}}\nonumber
\eea
and the odd quasinormal mode is
\bea
H_{o}(x)&=&\cosh^{\sigma+1}(\epsilon x) \sinh(\epsilon x)\ {}_2F_1\left(\frac{\sqrt{17}}{2},\frac{3}{2};\frac{3}{2};-\sinh^2(\epsilon x)\right)\label{dispe}\\
&=&\left[\sech(\epsilon x)\right]^{\frac{\sqrt{17}-1}{2}}\sinh(\epsilon x).\nonumber
\eea 

%Note that in both cases, at large $|x|$ and small $\epsilon$, Eq.~(\ref{g3}) becomes
%\beq
%(8m^2+\partial_{x}^2)G_3(x)=0
%\eeq
%and so $G_3$ tends to a plane wave.  Therefore, these apparent shape modes are, in fact, at subleading order, revealed to be continuum modes.  The $G(x)$ component in isolation would leak out of the Q-ball and so represents a Feshbach-type quasinormal mode.

\section{Numerical Results} \label{numsez}
\subsection{Our Results}

\begin{figure}
    \centering
    \includegraphics[width=1\linewidth]{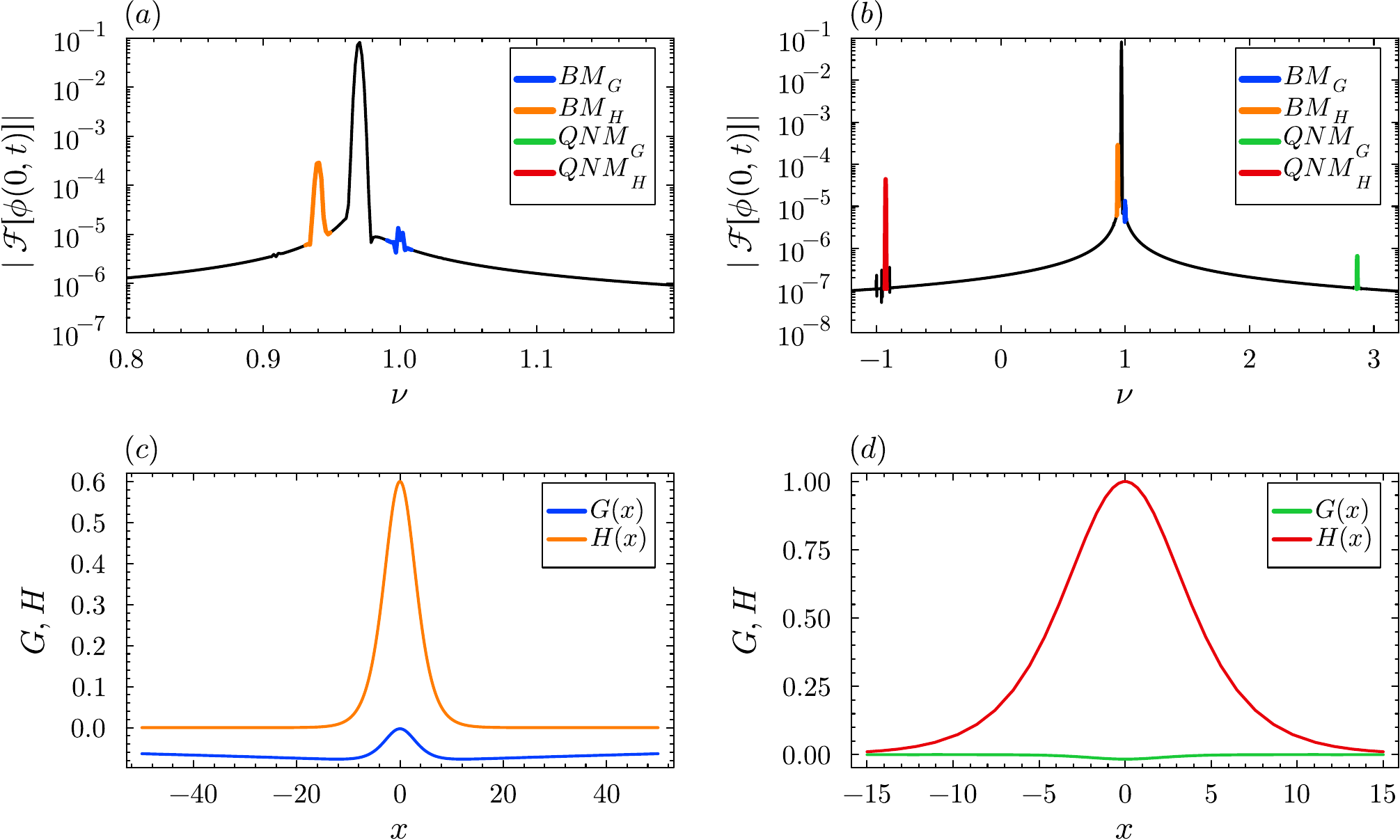}
    \caption{We set $m=1$.  (a),(b) Power spectrum of the field at the center for the squashed Q-ball $\Omega=0.97$ in the inverted $\phi^4$ model together with (c) the bound mode profile and (d) the even quasinormal mode profile.  Note that the bound mode extends far beyond the Q-ball itself.% I hope I did not messed up $H\leftrightarrow G$. These are even modes (squashing is symmetric). Jarah also considers odd+even.
    }
    \label{fig:FFT_Modes}
\end{figure}

In this section we set $m=1$.  In the case $\Omega=0.97$, we have shown the power spectrum of an even relaxing Q-ball in the top panels of Fig.~\ref{fig:FFT_Modes}.  Here the bound and the even quasinormal mode discussed in the text are evident.  Their shapes at this value of $\Omega$ are shown in the bottom panels.  The orange and blue peaks and curves  are respectively the $H$ and $G$ components of the bound corotating mode discussed in Subsec.~\ref{boundsez}.  It is evident in panel (c) that the blue curve extends far beyond the nominal size of the Q-ball, reflecting the fact that it is loosely bound.  The green and red peaks and curves are respectively the $G$ and $H$ terms in Eq.~(\ref{g3rot}), evaluated for the values of $\omega$ of the quasinormal modes in Eqs.~(\ref{qnmw}).  Note that, as expected, the red peak is much larger than the green peak, which appears at subleading order in the $\epsilon$ expansion.

The bound mode is found numerically to be at $\omega=0.029986$, compared to the leading $\epsilon$ expansion for the location of the threshold
\beq
\frac{\epsilon^2}{2}=0.0296.
\eeq
The disagreement is less than $\epsilon^4$.  The even quasinormal mode is found to be at $\Omega\pm 1.895794$ corresponding to a bound component at $-0.925794$ and an unbound component at $2.86579$.  This can be compared with our leading order results obtained by inserting Eq.~(\ref{qnmw}) into Eq.~(\ref{g3rot}), which for the bound component yields
\beq
-\Omega-\hat\omega_{2,e}\epsilon^2=-0.97+\left(\frac{7-\sqrt{17}}{4}\right)\left(1-0.97^2\right)\sim -0.927
\eeq
and for the unbound component
\beq
3\Omega+\hat\omega_{2,e}\epsilon^2=3*0.97+\left(\frac{\sqrt{17}-7}{4}\right)\left(1-0.97^2\right)\sim 2.867.
\eeq
The errors are of order $O(\epsilon^4)$, as expected as we have not included the $O(\epsilon^4)$ corrections in our expansion.

% In case of a BM $G\approx\sech(\epsilon x)^2$ and in case of a QNM $H\approx\sech(\epsilon x)^{3/2}$

\begin{figure}
    %\centering    \includegraphics[width=0.5\linewidth]{Odd_QNM-cropv2.pdf}
    \centering    \includegraphics[width=0.5\linewidth]{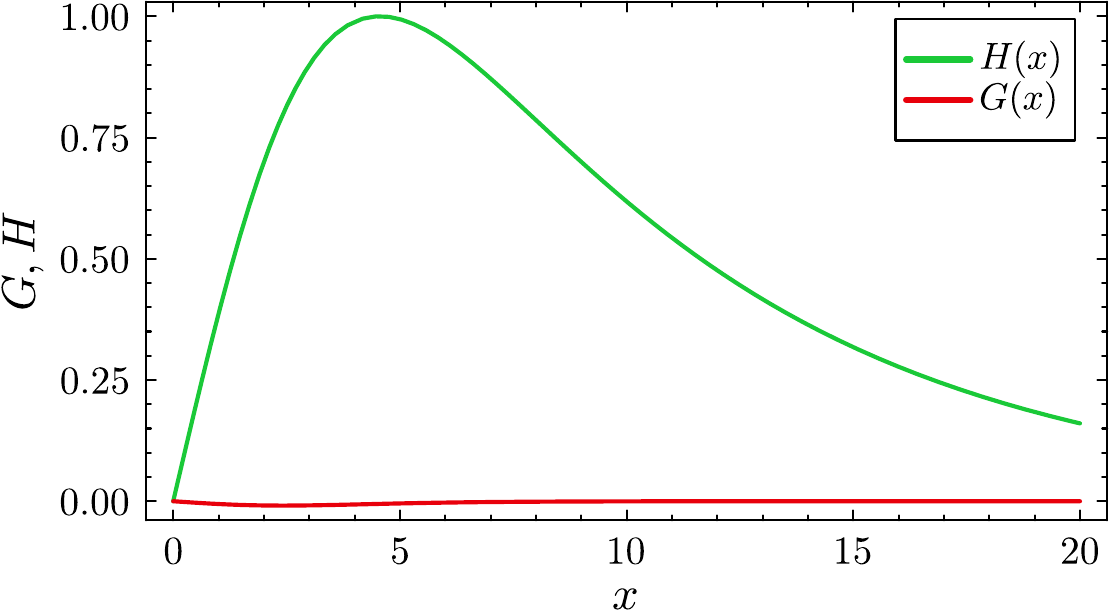}
    \caption{The odd quasinormal mode, evaluated numerically at $\Omega=0.97$.}
    \label{disfig}
\end{figure}

In Fig.~\ref{disfig} we provide a numerical evaluation of the odd quasinormal mode, again at $\Omega=0.97$.  $H(x)$ seen in the figure is a good fit to our leading order solution in Eq.~(\ref{dispe}).  Numerically, the bound and unbound components $H$ and $G$ appear at frequencies $-0.99059$ and $2.93059$ respectively.  These can be compared with our leading order results obtained by inserting Eq.~(\ref{qnmw}) into Eq.~(\ref{g3rot}), which for the bound component yields
\beq
-\Omega-\hat\omega_{2,o}\epsilon^2=-0.97+\left(\frac{11-3\sqrt{17}}{4}\right)\left(1-0.97^2\right)\sim -0.9902
\eeq
and for the unbound component
\beq
3\Omega+\hat\omega_{2,o}\epsilon^2=3*0.97+\left(\frac{3\sqrt{17}-11}{4}\right)\left(1-0.97^2\right)\sim 2.9302.
\eeq
In other words, the frequencies agree to well within $\epsilon^4$.

\begin{figure}
    \centering
    \includegraphics[width=0.8\linewidth]{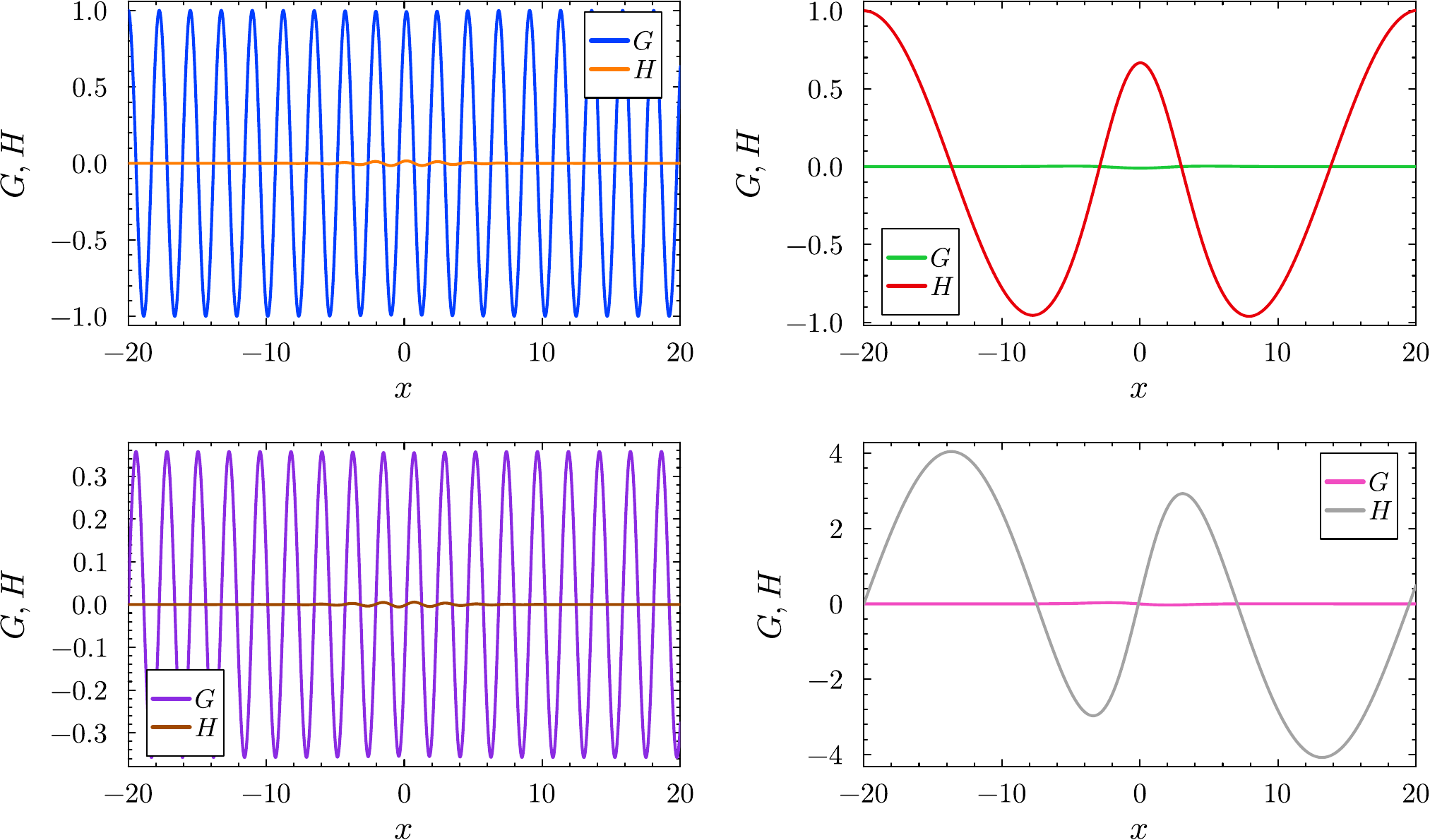}
    \caption{Four linearly independent solutions for $\Omega=0.97$, $\omega=2$, which is in the counterrotating regime.  Note that, due to the high wavenumber of $G$, the two components are nearly decoupled.}
    \label{fig:placeholder}
\end{figure}

In Fig.~\ref{fig:placeholder} we provide examples of even and odd continuum counterrotating excitations.

\begin{table}
\begin{center}
\begin{tabular}{|l|l|l|l|l|}
\hline
Mode&Analytic&Numerical&Ref.~\cite{qpert24}&Ref.~\cite{qpert08}\\
\hline\hline
Bound Mode $G$ ($\Omega=0.97$)&0.9996&0.999986&&\\
\hline
Bound Mode $H$ ($\Omega=0.97$)&0.9404&0.940014&&\\
\hline
Quasinormal Even Mode $H$ ($\Omega=0.97$)&-0.927&-0.925794&&\\
\hline
Quasinormal Even Mode $G$ ($\Omega=0.97$)&2.867&2.86579&&\\
\hline
Quasinormal Odd Mode $H$ ($\Omega=0.97$)&-0.9902&-0.99059
&&\\
\hline
Quasinormal Odd Mode $G$ ($\Omega=0.97$)&2.9302&2.93059&&\\
\hline
Bound Mode $G$ ($\Omega=\sqrt{3}/2$)&0.991&&0.9977&0.9980\\
\hline
Bound Mode $H$ ($\Omega=\sqrt{3}/2$)&0.741&&0.7343&0.7340\\
\hline
Quasinormal Even Mode $H$ ($\Omega=\sqrt{3}/2$)&$-0.686$&&-0.650&-0.645\\
\hline
Quasinormal Even Mode $G$ ($\Omega=\sqrt{3}/2$)&2.418&&2.383&2.3820\\
\hline
\end{tabular}
\end{center}
\caption{Frequencies of discrete modes} \label{numtab}
\end{table}

\subsection{Discrete Modes in the Literature}

Discrete modes in Q-balls have been reported previously in the literature.  In the Appendix of Ref.~\cite{qpert24}, in the case $\epsilon=0.5$ corresponding to $\Omega=\sqrt{3}/2$, the authors present a power spectrum with various peaks that they interpret.  They report bound states at frequencies of $0.9977$ and $0.7343$.  The first is just beneath the mass threshold and so is our loosely bound state from Subsec.~\ref{boundsez}.  This suggests that at $\epsilon=0.5$ the gap between the threshold and the bound state is $0.0023$.  Our argument then says that the other bound state should be at $2\Omega-0.9977\sim 0.7344$ which is a good fit.  They also find quasinormal mode peaks at frequencies of $-0.650$ and $2.383$ whereas, for the even quasinormal mode, we would expect the bound component to be at
\beq
-\Omega-\hat\omega_{2,e}\epsilon^2=-\frac{\sqrt{3}}{2}+\left(\frac{7-\sqrt{17}}{4}\right)\left(\frac{1}{4}\right)\sim-0.686
\eeq
and the unbound component at
\beq
3\Omega+\hat\omega_{2,e}\epsilon^2=3*\frac{\sqrt{3}}{2}+\left(\frac{\sqrt{17}-7}{4}\right)\left(\frac{1}{4}\right)\sim 2.418.
\eeq
Even at such a large value of $\epsilon$, our approximation works within about five percent, which is again of order $O(\epsilon^4)$.  We conclude that the bound and quasinormal modes found numerically in Ref.~\cite{qpert24} are indeed the same as those found in the present work.

Bound states in fact have been reported even earlier in Ref.~\cite{qpert08}.  Here, also at $\epsilon=0.5$, beating was found at a frequency of $0.132$, which is a good match for bound mode found above whose frequency with respect to the Q-ball is $0.9977-\sqrt{3}/2\sim 0.1317$.  The authors interpret it as a superposition of a Q-ball and a frequency one breather solution in an integrable deformation of this model.  Perhaps the deformation that the authors describe transforms our bound state into the breather of the deformed model.

The same paper finds a second discrete mode yielding a beating at a frequency of 1.516.  In other words, one expects excitations at $\sqrt{3}/2\pm 1.516$ corresponding to $-0.6450$ and $2.3820$.  These are therefore even quasinormal modes which were also identified in Ref.~\cite{qpert24}.  However, Ref.~\cite{qpert24} was the first to identify this excitation as a quasinormal mode rather than a bound mode.

These modes are all summarized in Table \ref{numtab}.

\section{What Next?}

Here, a systematic search has led to what appears to be a complete classification of the discrete modes of the Q-ball, including the modes that had been previously discovered numerically plus an apparently new, odd quasinormal mode.  The same approach could applied to other systems, such as the bright soliton of Ref.~\cite{kovtun}, which discovered a mysterious continuum pole that could indicate a quasinormal mode.

%The tower of $e^{-i(2n+1)\Omega t}$ dependences of the counterrotating modes is reminiscent of the tower of higher frequencies of the oscillon.  In that case, a Borel resummation shows that there is in fact a nonperturbative term outside of the tower, which indicates the oscillon's instability \cite{borel08}.  In the present case, we have found a tower of higher harmonics not in the Q-ball solution itself, but rather in the perturbations.  It would be interesting to again perform a Borel resummation and to see what interesting physics results.  For example, perhaps, as in the case of the oscillon solution, in addition to the periodic evolution, the counterrotating modes also enjoy a secular evolution which converts them into corotating modes, as in Refs.~\cite{qsuper1,qsuper2,qsuper3}.  This would imply that the monodromy matrix describing the time evolution by one period of the Floquet modes is not diagonalizable, but rather is upper triangular, suggesting that the Floquet Hamiltonian itself is not Hermitian.  The usual construction of the Q-ball ground state as a tensor product of Floquet mode ground states would then fail, opening the possibility that the Q-ball is not stable, as a result of the effect in the above references.  
%the counterrotating modes are never perfectly Floquet, but have some interesting monodromy that affects their secular evolution, and so potentially the quantum Q-ball.

What about quantum Q-balls?  It has long been appreciated \cite{dhn2} that the first step in quantizing a classical field theory solution is to find its linearized perturbations.  In the case of a time-independent solution these are normal modes.  

The leading quantum correction to the energy of a periodic classical solution may be obtained \cite{dhnsg} using Floquet modes, which are periodic up to a phase.  The Floquet modes do not span the space of deformations.  On the contrary, some linearized perturbations, such as the infinitesimal boost or the amplitude shift, have more general monodromies.  In order to decompose the quantum field, one needs a complete basis of perturbations.  As a result, such modes are also required in order to understand the dynamics of the quantum theory, even if they are not needed to calculate the one-loop correction to the energy.

In the present note, we have presented the linearized perturbations of the Q-ball at leading order in the Q-ball's amplitude.  In the case of the oscillon, this allowed a construction of the quantum ground state in Ref.~\cite{qosc25}.  With the ground state in hand, one can calculate the spontaneous and induced rate for radiation emission, scattering amplitudes, and more.  

The natural next step is to follow the procedure used in that work to find the ground state of the Q-ball.  Although the Q-ball is not integrable, it will be possible to test our results, for example by comparing the derived stress tensor with that obtained using the methods of Ref.~\cite{stress26}. Classically, the Q-ball is stable in isolation, but exhibits induced emission \cite{qsuper1,qsuper2,qsuper3}.  The key question, which may be answered once the Q-ball is quantized, is whether the quantum Q-ball spontaneously radiates and so is unstable even in isolation, potentially yielding an even richer dark matter phenomenology, as described in Refs.~\cite{dm0,dm1,dm2, Weinberg, SS, Strigari, Peter, Fuss}. One might object that stability of the low amplitude Q-ball is guaranteed by the conservation of charge and convexity of the mass-charge relation.  However, the approximate Floquet oscillator ground states lead to an IR-divergent zero-point energy, which may in principle compensate for the energy lost when the charge is radiated.

%The fact that Q-ball might be unstable does not necessary mean that they are phenomenologically not interesting.
%In contrast, although dark matter is usually assumed to be stable, 
%there exist a number of extensions which take into account the possible decay into lighter particles \cite{}.

There is also an intriguing possibility that modes of oscillons can originate in the spectral structure of a certain $Q$-ball solutions. This idea is based on the recently established renormalization-group-inspired relation between oscillons and $Q$-balls \cite{Blaschke:2024dlt}, where a generic, not necessary small amplitude oscillon is sourced in a single or two-$Q$-ball solution of a universal complex field theory. 

\section*{Acknowledgement}

\noindent
This work was supported by the Higher Education and Science Committee of the Republic of Armenia (Research Project Nos. 24PostDoc/2‐1C009 and 24RL-1C047). This work was also partly supported by a short term scientific mission grant from the COST action CA22113 THEORY-CHALLENGES. Y.~S. would like to thank E.~Kim and E.~Nugaev for inspiring and valuable discussions. 
Y.~S. is partially supported by the FCT mobility programme, grant RE-C06-i06.M02. 
H.~L. acknowledges partial support from program of collaboration 
between JINR and the Republic of Armenia. A. W. acknowledges the support from the Spanish Ministerio de Ciencia e Innovacion (MCIN) with funding from the grant PID2023-148409NB-I00 MTM.

\end{document}